\documentclass[notitlepage,reqno,a4paper]{article}
\usepackage{}
\usepackage{amsfonts}
\usepackage{subfigure}
\usepackage{cite}
\usepackage{mathrsfs}
\usepackage{stmaryrd}
\usepackage{amsmath}
\usepackage{amssymb}
\usepackage{graphicx}
\usepackage{verbatim}
\usepackage{indentfirst}
\usepackage{txfonts}

\begin{document}
\begin{center}\large \textbf{Fractional Green's function for the time-dependent scattering problem in the Space-time-fractional quantum mechanics}
\end{center}
\begin{center}\small {Jianping Dong\footnote{Email:Dongjp.sdu@gmail.com}}\\
{\itshape Department of Mathematics, College of Science, Nanjing
University of Aeronautics and Astronautics, Nanjing 210016,
China}\end{center}
% ----------------------------------------------------------------
\begin{quote}  Integral form of the space-time-fractional Schr\"odinger equation for the scattering
problem in the fractional quantum mechanics is studied in this paper. We define the fractional Green's function for the space-time fractional Schr\"odinger equation and express it in terms of Fox's H-function and in a computable series form.  The asymptotic formula of the Green's function for large argument is also obtained, and
applied  to study the fractional quantum scattering problem. We get the approximate scattering wave function with correction of every order. \end{quote}
% ----------------------------------------------------------------
\section{\uppercase{Introduction}}
 The fractional calculus \cite{pod,kilbas}, which is a generalization to the standard (integer) one, has been successfully
applied in many science and engineering fields, such as  anomalous transport, viscoelastic material,
signal analysis and processing, control system \cite{xu,west,Tarasov,Monje}. In recent years, the fractional calculus enters the world of quantum mechanics.
In the fractional quantum mechanics \cite{laskin1,laskin2}, the master equation is the fractional Schr\"odinger equation \cite{laskin3} instead of the standard one.
 The standard Schr\"odinger equation \cite{levin,grif} was reformulated by Feynman and Hibbs \cite{feynman} using the path integral
 approach considering the Gaussian probability distribution. The L\'evy stochastic process is a natural generalization of the Gaussian process.
The possibility of developing the path integral over the paths of the L\'evy motion was
discussed by Kac \cite{kac}, who pointed out that the L\'evy path
integral generates the functional measure in the space of left (or
right) continued functions having only discontinuities of the first
kind. Then, Laskin\cite{laskin1} generalized
Feynman path integral to L\'evy one, and developed a
space-fractional Schr\"odinger equation containing the Riesz
fractional derivative \cite{laskin3,kilbas}. Then, he constructed
the fractional quantum mechanics and showed some properties of the
space fractional quantum system \cite{laskin2,laskin3,laskin4,laskin5}. \par Afterwards, Naber\cite{naber} constructed
a time-fractional Schr\"odinger equation by introducing the Caputo fractional derivative \cite{pod}
instead of the first order derivative over time to the standard
Schr\"odinger equation to describe non-Markovian evolution in
quantum physics.  The Hamiltonian for the time fractional quantum system was
found to be non-Hermitian and not local in time. Naber solved the
time fractional Schr\"odinger equation for a free particle and for a
potential well. Probability and the resulting energy levels are
found to increase over time to a limiting value depending on the
order of the time derivative. More recently, from the standard Schr\"odinger equation, Wang and Xu \cite{wang1}
established a Schr\"odinger equation with both space and time
fractional derivatives, and solved the generalized Schr\"odinger
equation for a free particle and for an infinite rectangular
potential well. Then, a similar space-time-fractional Schr\"odinger
equation is obtained by Dong and Xu \cite{dong3} from the space-fractional Schr\"odinger equation. They expressed this fractional
equation in a more simple form, and studied the time evolution behaviors of the space-time-fractional quantum system in the
time-independent potential fields.
\par At present, the progresses for the space-time-fractional quantum system is fewer. Besides the results given in Refs.~\cite{wang1,dong3} mentioned before, Jiang \cite{jiang} developed a time-space fractional Schr\"odinger equation containing a nonlocal term, and obtained the time dependent solutions in terms of the H-function. All of these results are in the one-dimensional case. This paper focuses on the time-dependent scattering problem in the fractional quantum system described by the space-time-fractional Schr\"odinger equation, given by Dong and Xu, in the three-dimensional case.  We will define the Green's function for the time-dependent 3D space-time-fractional Schr\"odinger equation for the scattering problem, and the Green's function will be  calculated in terms of Fox's H-function and in a computable series form. The asymptotic property of the Green's function will also be discussed and applied to the fractional scattering problem.

\section{\uppercase{Green's Function of the Fractional Schr\"odinger
Equation}} \label{sec1}The space-time fractional Schr\"odinger equation
\cite{laskin3} obtained by Laskin reads (in two dimensions)
\begin{equation} (i\hslash)^\beta
D^\beta_t\psi(\mathrm{\textbf{r}},t)=\mathscr{H}_\alpha\psi(\mathrm{\textbf{r}},t), \label{fse1}
\end{equation}
where $\psi(\mathrm{\textbf{r}},t)$ is the time-dependent wave
function , and
\begin{equation}\mathscr{H}_\alpha=\mathcal {M}\mbox{ }[-D_\alpha (\hslash\nabla)^{\alpha}+V(\mathrm{\textbf{r}},t)].\label{fse2}
\end{equation}
Here, $$\mathcal {M}=\frac{\hslash^\beta}{E_pT_p^\beta}$$, $D_{\alpha}$ with physical dimension
$[D_{\alpha}]=\text{[Energy]}^{1-\alpha}\times
\text{[Length]}^\alpha \times \text{[Time]}^{-\alpha}$ is dependent
on $\alpha$ [we have $D_{\alpha}=1/2m$ for $\alpha=2$, $m$ denotes the mass
of a particle] and $(\hslash\nabla)^{\alpha}$ is the quantum Riesz
fractional operator \cite{kilbas,laskin1} defined by
\begin{equation}
(\hslash\nabla)^{\alpha}\psi(\mathrm{\textbf{r}},t)=-\frac1{(2\pi\hslash)^2}\int\mathrm{d}^3\mathrm{\textbf{p}}\text{e}^
{i\mathrm{\textbf{p}}\cdot\mathrm{\textbf{r}}/\hslash}|\mathrm{\textbf{p}}|^\alpha\int\text{e}^
{-i\mathrm{\textbf{p}}\cdot\mathrm{\textbf{r}}/\hslash}\psi(\mathrm{\textbf{r}},t)\mathrm{d}^3\mathrm{\textbf{r}}
\label{riesz1}.
\end{equation}
Note that by use of the method of dimensional analysis we have given
a specific expression of $D_{\alpha}$ in Ref.~\cite{dong3} as
$D_{\alpha}=\bar{c}^{2-\alpha}/(\alpha m^{\alpha-1})$, where
$\bar{c}$ denotes the characteristic velocity of the
non-relativistic quantum system.

 Now, we define a Green's function of the FSE by
\begin{equation}
[(i\hslash)^\beta
D^\beta_t+\mathcal {D}
(\hslash\nabla)^{\alpha}]G(\mathrm{\textbf{r}},t;\mathrm{\textbf{r}}^\prime,t^\prime)
=\delta(\textbf{r}-\mathrm{\textbf{r}}^\prime)\delta(t-t^\prime),\label{green}
\end{equation}with the causality condition
\begin{equation}
G(\mathrm{\textbf{r}},t;\mathrm{\textbf{r}}^\prime,t^\prime)=0,\quad\mbox{when
} t<t^\prime. \label{causality}
\end{equation} Here, $\mathcal {D}=\mathcal {M}\cdot D_\alpha$. Then, the space-FSE~(\ref{fse1}) becomes an integral equation,
\begin{equation}
\psi(\mathrm{\textbf{r}},t)=\psi_0(\mathrm{\textbf{r}},t)+\mathcal {D}\iint
G(\mathrm{\textbf{r}},t;\mathrm{\textbf{r}}^\prime,t^\prime)V(\mathrm{\textbf{r}}^\prime,t^\prime)\psi(\mathrm{\textbf{r}}^\prime,t^\prime)
\mathrm{d}^3\mathrm{\textbf{r}}^\prime\mathrm{d}t^\prime,\label{integralform}
\end{equation} in which $\psi_0(\mathrm{\textbf{r}},t)$ satisfies the free-particle
Schr\"odinger equation,
\begin{equation}
[i\hslash\frac{\partial}{\partial t}+\mathcal {D}
(\hslash\nabla)^{\alpha}]\psi_0(\mathrm{\textbf{r}},t)=0.\label{freeparticle}
\end{equation}By use of the method of separation of variables, the basic solution
to Eq.~(\ref{freeparticle}) can be easily obtained.
\begin{equation}
\psi_0(\mathrm{\textbf{r}},t)=\mbox{e}^{i(\mathrm{\textbf{k}}\cdot\mathrm{\textbf{r}}-Et)/\hslash},\text{
{} {} (a constant product factor is omitted )}\label{freeparticle1}
\end{equation}
where $E$ denotes the energy of the free particle, and
$\mathrm{\textbf{k}}=(k_x,k_y), $ in which $k_x,k_y$ are arbitrary
constants but satisfying
$|\mathrm{\textbf{k}}|=\sqrt{k_x^2+k_y^2}=(E/\mathcal {D})^{1/\alpha}$.
Replacing $\mathrm{\textbf{k}}$ by momentum $\mathrm{\textbf{p}}$,
and $E$ by $\mathcal {D}|\mathrm{\textbf{p}}|^\alpha$ respectively, the
free particle solution $\psi_0(\mathrm{\textbf{r}},t)$ can be
changed to the fractional plane wave solution \cite{laskin4},
\begin{equation}\psi_0(\mathrm{\textbf{r}},t)=\mbox{e}^{i(\mathrm{\textbf{p}}\cdot\mathrm{\textbf{r}}-\mathcal {D}|\mathrm{\textbf{p}}|^\alpha
t)/\hslash}.\label{fracplane}
\end{equation}
\\ Now we turn back to solve Eq.~(\ref{green}).  Defining the
Fourier-Laplace transform pair, with respect to $\mathrm{\textbf{r}}$ and
$t$, of the Green's function
$G(\mathrm{\textbf{r}},t;\mathrm{\textbf{r}}^\prime,t^\prime)$ as
\begin{align}
&\hat{G}(\mathrm{\textbf{p}},s;\mathrm{\textbf{r}}^\prime,t^\prime)
=\int\mathrm{d}^3\mathrm{\textbf{r}}\int_{0}^{+\infty}\mathrm{d}t\text{e}^{-i\mathrm{\textbf{p}}\cdot\mathrm{\textbf{r}}/\hslash-\omega
t}G(\mathrm{\textbf{r}},t;\mathrm{\textbf{r}}^\prime,t^\prime),\label{fourier1}
\\&G(\mathrm{\textbf{r}},t;\mathrm{\textbf{r}}^\prime,t^\prime)
=\int\frac{\mathrm{d}^3\mathrm{\textbf{p}}}{(2\pi\hslash)^3}\int_{c-i\infty}^{c+i\infty}\frac{\mathrm{d}s}{2\pi i}\text{e}^{i\mathrm{\textbf{p}}\cdot\mathrm{\textbf{r}}/\hslash+s
t}\hat{G}(\mathrm{\textbf{p}},s;\mathrm{\textbf{r}}^\prime,t^\prime).\label{fourier2}
\end{align}
After taking the above Fourier-Laplace transform with respect to $\mathrm{\textbf{r}}$ and
$t$, Eq.~(\ref{green}) can be changed to
\begin{equation}
 (i\hslash)^\beta[s^\beta \hat{G}(\mathrm{\textbf{p}},s;\mathrm{\textbf{r}}^\prime,t^\prime)+s^{\beta-1}\bar{G}(\mathrm{\textbf{p}},0;\mathrm{\textbf{r}}^\prime,t^\prime)]-\mathcal {D}|\mathrm{\textbf{p}}|^\alpha\hat{G}(\mathrm{\textbf{p}},\omega;\mathrm{\textbf{r}}^\prime,t^\prime)
=\text{e}^{-i\mathrm{\textbf{p}}\cdot\mathrm{\textbf{r}^\prime}/\hslash-s
t^\prime},
\end{equation}in which $\bar{G}(\mathrm{\textbf{p}},0;\mathrm{\textbf{r}}^\prime,t^\prime)$ is the Fourier transform of $G(\mathrm{\textbf{r}},0;\mathrm{\textbf{r}}^\prime,t^\prime)$ with respect to $\mathrm{\textbf{r}}$. Then, taking account of the causality condition given  by (\ref{causality}), we have $G(\mathrm{\textbf{r}},0;\mathrm{\textbf{r}}^\prime,t^\prime)=0$, so that $\bar{G}(\mathrm{\textbf{p}},0;\mathrm{\textbf{r}}^\prime,t^\prime)=0$. Now, we can obtain
\begin{equation}
\hat{G}(\mathrm{\textbf{p}},\omega;\mathrm{\textbf{r}}^\prime,t^\prime)=\frac{\text{e}^{-i\mathrm{\textbf{p}}\cdot\mathrm{\textbf{r}^\prime}/\hslash}\text{e}^{-s
t^\prime}}{(i\hslash)^\beta s^\beta-\mathcal {D}|\mathrm{\textbf{p}}|^\alpha}.
\end{equation}
Inverting the Fourier-Laplace transform gives
\begin{equation}
G(\mathrm{\textbf{r}},t;\mathrm{\textbf{r}}^\prime,t^\prime)=\int\frac{\mathrm{d}^3\mathrm{\textbf{p}}}{(2\pi\hslash)^3}
\text{e}^{i\mathrm{\textbf{p}}\cdot(\mathrm{\textbf{r}}-\mathrm{\textbf{r}^\prime})/\hslash}
\int_{c-i\infty}^{c+i\infty}\frac{\mathrm{d}s}{2\pi i}\frac{\text{e}^{s(t-
t^\prime)}}{(i\hslash)^\beta s^\beta-\mathcal {D}|\mathrm{\textbf{p}}|^\alpha}.\label{greenint}
\end{equation} To calculate the integrals in the above formula, more
work is needed.  Firstly, the $s$ integration can be calculated by use of the residue theory,  we have
\begin{equation}
\frac{1}{2\pi i}\int_{c-i\infty}^{c+i\infty}\frac{\text{e}^{s(t-
t^\prime)}}{(i\hslash)^\beta s^\beta-\mathcal {D}|\mathrm{\textbf{p}}|^\alpha}\mathrm{d}s=\mbox{Res}\left\{\frac{\text{e}^{s(t-
t^\prime)}}{(i\hslash)^\beta s^\beta-\mathcal {D}|\mathrm{\textbf{p}}|^\alpha},s,s_0\right\}=\frac{\text{e}^{s_0(t-
t^\prime)}}{\beta(i\hslash)^\beta s_0^{\beta-1}}=
\frac{\text{e}^{-i|\mathrm{\textbf{p}}|^{\alpha/\beta}\mathcal {D}^{1/\beta}(t-t^\prime)/\hslash}}
{i\hslash\beta\mathcal {D}^{(\beta-1)/\beta}}|\mathrm{\textbf{p}}|^{-\alpha(\beta-1)/\beta}.\label{calfors}
\end{equation}
Here, $\mbox{Res}\left\{*,s,s_0\right\}$ denotes the residue \cite{markj} of $*$, with respect to $s$, at the unique pole $s_0$, and $s_0=(\mathcal {D}|\mathrm{\textbf{p}}|^\alpha)^{1/\beta}/(ih)$. Now Eq.~(\ref{greenint}) becomes
\begin{equation}
G(\mathrm{\textbf{r}},t;\mathrm{\textbf{r}}^\prime,t^\prime)=\frac{N_1}{(2\pi\hslash)^3}
\int\text{e}^{i\mathrm{\textbf{p}}\cdot(\mathrm{\textbf{r}}-\mathrm{\textbf{r}^\prime})/\hslash}
\text{e}^{-i\xi|\mathrm{\textbf{p}}|^\nu}|\mathrm{\textbf{p}}|^{-\gamma}\mathrm{d}^3\mathrm{\textbf{p}},\quad
t>t^\prime,\label{greenint1}
\end{equation}in which $$N_1=\frac{1}
{i\hslash\beta\mathcal {D}^{(\beta-1)/\beta}},\xi=\mathcal {D}^{1/\beta}(t-t^\prime)/\hslash,\nu=\alpha/\beta,\gamma=\alpha(\beta-1)/\beta.$$
To execute the above integration, we choose the spherical
coordinates $(p,\theta,\varphi)$, with the positive direction of the
$p$-axis along $\mathrm{\textbf{r}}-\mathrm{\textbf{r}}^\prime$.
Then,
$\mathrm{\textbf{p}}\cdot(\mathrm{\textbf{r}}-\mathrm{\textbf{r}^\prime})=p|\mathrm{\textbf{r}}-\mathrm{\textbf{r}^\prime}|\cos\theta$,
in which $p$ and $|\mathrm{\textbf{r}}-\mathrm{\textbf{r}}^\prime|$
denote the magnitudes of the vectors $\mathrm{\textbf{p}}$ and
$\mathrm{\textbf{r}}-\mathrm{\textbf{r}}^\prime$, respectively.
Thus, Eq.~(\ref{greenint1}) is converted into
\begin{equation}
\begin{split}
G(\mathrm{\textbf{r}},t;\mathrm{\textbf{r}}^\prime,t^\prime)&=\frac{N_1}{(2\pi\hslash)^3}\int^{\pi} _{0}\mathrm{d}\theta\int^{2\pi}
_{0}\mathrm{d}\varphi\int^{+\infty}
_{0}\text{e}^{ip|\mathrm{\textbf{r}}-\mathrm{\textbf{r}^\prime}|\cos\theta/\hslash}\text{e}^{-i\xi|\mathrm{\textbf{p}}|^\nu}|\mathrm{\textbf{p}}|^{2-\gamma}
\sin\theta\mathrm{d}p
\\&=\frac{N_1}{2\pi^2\hslash^2|\mathrm{\textbf{r}}-\mathrm{\textbf{r}^\prime}|}\int^{+\infty}
_{0}p^{1-\gamma}\sin(p|\mathrm{\textbf{r}}-\mathrm{\textbf{r}^\prime}|/\hslash)\text{e}^{-i\xi p^\nu}\mathrm{d}p,\label{greenint2}
\end{split}
\end{equation}In the last integral, substituting $p$ for $\xi^{1/\nu}p$ gives
\begin{equation}
G(\mathrm{\textbf{r}},t;\mathrm{\textbf{r}}^\prime,t^\prime)=N\cdot\frac{I(x)}{|\mathrm{\textbf{r}}-\mathrm{\textbf{r}^\prime}|}
,\label{greenint3}
\end{equation}
in which \begin{equation}
I(x)=\int^{+\infty}_{0}p^{1-\gamma}\sin(px)\text{e}^{-ip^\nu}\mathrm{d}p,\quad
x\geq0, \label{greenintir}
\end{equation}
\begin{equation}
x=\frac{|\mathrm{\textbf{r}}-\mathrm{\textbf{r}^\prime}|}{\xi^{1/\nu}\hslash},\mbox{ and } N=\frac{N_1\xi^{(\gamma-2)/\nu}}{2\pi^2\hslash^2}.\label{xxx}
\end{equation}
Using the Mellin transform \cite{Polyanin} and its inverse
transform, $I(x)$ can be expressed in terms of Fox's $H$-function.
Taking Mellin transform  to $I(x)$ with respect to $x$ yields
\begin{equation}
\begin{split} \tilde{I}(s)&=\mathcal {M}\text{ }\Big\{I(x),s\Big\}=\int^{+\infty}_{0}p^{1-\gamma}\text{e}^{-ip^\nu}\left(\int^{+\infty}_{0}\sin(px)x^{s-1}\mathrm{d}x\right)\mathrm{d}p
\\&=\Gamma(s)\sin(\frac{\pi s}{2})\int^{+\infty}_{0}p^{1-\gamma-s}\text{e}^{-ip^\nu}\mathrm{d}p,\quad(\mbox{Re
}s<1).
\end{split}\label{greenintirmellin}
\end{equation} \begin{figure}[h]
   \centering
 \includegraphics[width=6cm]{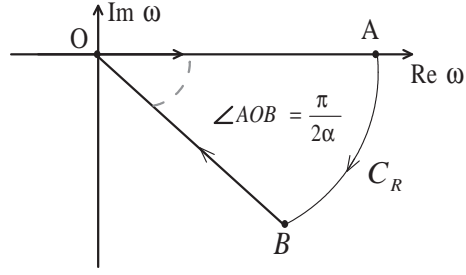}
 \caption{The contour used to calculate the last integral in
 Eq.~(\ref{greenintirmellin}).}
 \label{fig3}
\end{figure}Then using the contour in Fig. \ref{fig3}, the above integral can be
calculated,
\begin{equation}
\int^{+\infty}_{0}p^{1-\gamma-s}\text{e}^{-ip^\nu}\mathrm{d}p=\frac1\nu\Gamma(\frac{2-\gamma-s}{\nu})\mbox{e}^{-\frac{2-s}{2\nu}\pi
i}.\label{greenintirmellin1}
\end{equation} Thus,
\begin{equation}
\tilde{I}(s)=\frac1\nu\Gamma(s)\Gamma(\frac{2-\gamma-s}{\nu})\sin(\frac{\pi
s}{2})\mbox{e}^{-\frac{2-\gamma-s}{2\nu}\pi
i}=\frac1\nu[\tilde{I}_1(s)-i\tilde{I}_2(s)],\label{greenintirmellinfinal}
\end{equation}where
\begin{align}
\begin{split}\tilde{I}_1(s)&=\Gamma(s)\Gamma(\frac{2-\gamma-s}{\nu})\sin(\frac{\pi
s}{2})\cos(\frac{2-\gamma-s}{2\nu}\pi)\\&=\frac{\pi^2\Gamma(s)\Gamma(\frac{2-\gamma-s}{\nu})}{\Gamma(\frac{s}{2})\Gamma(1-\frac{s}{2})\Gamma(\frac12+\frac{2-\gamma-s}{2-\gamma})
\Gamma(\frac12-\frac{2-\gamma-s}{2\nu})},
\end{split}\label{greenintirmellinfinal1}
\\\begin{split}\tilde{I}_2(s)&=\Gamma(s)\Gamma(\frac{2-\gamma-s}{\nu})\sin(\frac{\pi
s}{2})\sin(\frac{2-\gamma-s}{2\nu}\pi)\\&=\frac{\pi^2\Gamma(s)\Gamma(\frac{2-\gamma-s}{\nu})}{\Gamma(\frac{s}{2})\Gamma(1-\frac{s}{2})\Gamma(\frac{2-\gamma-s}{2\nu})
\Gamma(1-\frac{2-\gamma-s}{2\nu})}.
\end{split}\label{greenintirmellinfinal2}
\end{align}Note that the formulas \cite{Polyanin} $$\sin(\pi z)=\frac{\pi}{\Gamma(z)\Gamma(1-z)},\mbox{ and }\cos(\pi z)=\frac{\pi}{\Gamma(1/2+z)\Gamma(1/2-z)}$$ have been used here.
Inverting the Mellin transform and comparing the expression with the
definition of Fox's $H$-function (see Eqs.~(\ref{a1})and
(\ref{a1add}) in the Appendix of this paper or
Refs.~\cite{mathai,mathai0,kilbas0}), we obtain
\begin{equation}
\begin{split}I(x)=&\frac{1}{2\pi i}\int^{c+i\infty}
_{c-i\infty}\tilde{I}(s)x^{-s}\mathrm{d}s
\\=&\frac{\pi^2}{\nu}\Bigg\{H^{1,1}_{3,3}\left[x\text{
}\Bigg|
\begin{aligned}&{\text{
}(1-(2-\gamma)/\nu,1/\nu),(0,1/2),(1/2-(2-\gamma)/(2\nu),1/(2\nu))}
\\&{\text{
}(0,1),(0,1/2),(1/2-(2-\gamma)/(2\nu),1/(2\nu))}
\end{aligned}\right]-
\\&iH^{1,1}_{3,3}\left[x\text{
}\Bigg|
\begin{aligned}&{\text{
}(1-(2-\gamma)/\nu,1/\nu),(0,1/2),(1-(2-\gamma)/(2\nu),1/(2\nu))}
\\&{\text{
}(0,1),(0,1/2),(1-(2-\gamma)/(2\nu),1/(2\nu))}
\end{aligned}\right]\Bigg\}
\\=&\pi^2\left[H_1(x^\nu)-iH_2(x^\nu)\right],
\end{split}\label{greenintirfinal}
\end{equation}
where
\begin{align}
H_1(x)&=H^{1,1}_{3,3}\left[x\text{ }\Bigg|
\begin{aligned}&{\text{
}(1-(2-\gamma)/\nu,1),(0,\nu/2),(1/2-(2-\gamma)/(2\nu),1/2)}
\\&{\text{
}(0,\nu),(0,\nu/2),(1/2-(2-\gamma)/(2\nu),1/2)}
\end{aligned}\right],\label{hfun1}
\\H_2(x)&=H^{1,1}_{3,3}\left[x\text{ }\Bigg|
\begin{aligned}&{\text{
}(1-(2-\gamma)/\nu,1),(0,\nu/2),(1-(2-\gamma)/(2\nu),1/2)}
\\&{\text{
}(0,\nu),(0,\nu/2),(1-(2-\gamma)/(2\nu),1/2)}
\end{aligned}\right].\label{hfun2}
\end{align}Note that the first property of the $H$-function in the Appendix has been used (or see Property $1.4$ on page $12$ of Ref.~\cite{mathai0} ).
Finally, we get
\begin{equation}
\begin{split}
&G(\mathrm{\textbf{r}},t;\mathrm{\textbf{r}}^\prime,t^\prime)=\frac{N\pi^2}{|\mathrm{\textbf{r}}-\mathrm{\textbf{r}^\prime}|}\cdot
\left[H_1\left(\frac{|\mathrm{\textbf{r}}-\mathrm{\textbf{r}^\prime}|^{\nu}}{\xi\hslash^{\nu}}\right)
-iH_2\left(\frac{|\mathrm{\textbf{r}}-\mathrm{\textbf{r}^\prime}|^{\nu}}{\xi\hslash^{\nu}}\right)\right]
\\=&\frac{\mathcal {D}^{(\gamma+\nu-2)/(\beta\nu)-1}\hslash^{(2-\gamma)/\nu-3}}{2\beta(t-t^\prime)^{(2-\gamma)/\nu}|\mathrm{\textbf{r}}-\mathrm{\textbf{r}^\prime}|i}
\left[H_1\left(\frac{|\mathrm{\textbf{r}}-\mathrm{\textbf{r}^\prime}|^{\nu}}{\mathcal {D}^{1/\beta}(t-t^\prime)\hslash^{\nu-1}}\right)-
iH_2\left(\frac{|\mathrm{\textbf{r}}-\mathrm{\textbf{r}^\prime}|^{\nu}}{\mathcal {D}^{1/\beta}(t-t^\prime)\hslash^{\nu-1}}\right)\right],\quad
t>t^\prime.
\end{split}\label{greeninh}
\end{equation}

\section{\uppercase{Computable Series Form of the Green's Function}}
\label{sec3}Using the series form of the Fox's $H$-function (see
Property 3 in the Appendix of this paper or \S3.7 in
Ref.~\cite{mathai}), a computable form of the Green's Function can
be obtained. For the $H$-function $H_1(x)$ given by
Eq.~(\ref{hfun1}), we have
\begin{equation}
\begin{split}
H_1(x)&=\sum^{\infty}_{k=0}\frac{\Gamma((2-\gamma)/\nu+k/\nu)}{\Gamma(1+k/2)\Gamma(-k/2)\Gamma(1/2+(2-\gamma)/(2\nu)+k/(2\nu))\Gamma(1/2-(2-\gamma)/(2\nu)-k/(2\nu))}
\frac{(-1)^k}{k!}\frac{x^{k/\nu}}{\nu}
\\&=\sum^{\infty}_{k=0}\Gamma\left(\frac{2-\gamma+k}{\nu}\right)\frac{\sin(-k\pi/2)}{\pi}\frac{\cos[(1/\nu+(k-\gamma)/(2\nu))\pi]}{\pi}\frac{(-1)^k}{k!}\frac{x^{k/\nu}}{\nu}
\\&=\frac{1}{\nu\pi^2}\sum^{\infty}_{n=0}\Gamma\left(\frac{2n+3-\gamma}{\nu}\right)\cos\left[\frac{2n+3-\gamma}{2\nu}\pi\right]\frac{(-1)^{n}}{(2n+1)!}x^{(2n+1)/\nu}.\label{seriesforh1}
\end{split}
\end{equation}
In a similar way, we obtain
\begin{equation}
\begin{split}
H_2(x)=\frac{1}{\nu\pi^2}\sum^{\infty}_{n=0}\Gamma\left(\frac{2n+3-\gamma}{\nu}\right)\sin\left[\frac{2n+3-\gamma}{2\nu}\pi\right]\frac{(-1)^{n}}{(2n+1)!}x^{(2n+1)/\nu}.
\end{split}\label{seriesforh2}
\end{equation}
Then, the Green's Function given by Eq.~(\ref{greeninh}) can be
expressed in a series form as
\begin{equation}
\begin{split}
G(\mathrm{\textbf{r}},t;\mathrm{\textbf{r}}^\prime,t^\prime)=\frac{N}{\nu|\mathrm{\textbf{r}}-\mathrm{\textbf{r}^\prime}|}\cdot
\sum^{\infty}_{n=0}\Gamma\left(\frac{2n+3-\gamma}{\nu}\right)\exp\left(-\frac{2n+3-\gamma}{2\nu}\pi
i\right)\frac{(-1)^{n}}{(2n+1)!}\left(\frac{|\mathrm{\textbf{r}}-\mathrm{\textbf{r}^\prime}|}{\xi^{1/\nu}\hslash}\right)^{2n}.
\end{split}\label{greeninseries}
\end{equation}

 When $\beta=1$, this formula reduces to the space-fractional case, see Eq.() in Ref. .

\section{\uppercase{Asymptotic property of the Green's function and Its Application to the Scattering Problem}}
\label{sec4} \par In quantum mechanics, the scattering theory
\cite{levin,grif} is often used to study the inner structure of a
matter. The fractional quantum mechanics is a natural generalization
to the standard quantum mechanics, so the research on the
generalized quantum scattering problem under the framework of
fractional quantum mechanics is meaningful. In the scattering
problems, we usually consider the behavior of the particles far away
from the scattering center, and assume the potential
$V(\mathrm{\textbf{r}})$ is non-zero only in a small domain.
Therefore, for $\mathrm{\textbf{r}}$ and
$\mathrm{\textbf{r}}^\prime$ in Eq.~(\ref{integralform}), we have
$|\mathrm{\textbf{r}}|>>|\mathrm{\textbf{r}}^\prime|$, or
$|\mathrm{\textbf{r}}-\mathrm{\textbf{r}}^\prime|\rightarrow\infty$.
In this section, we study the asymptotic properties of the Green's
function when
$|\mathrm{\textbf{r}}-\mathrm{\textbf{r}}^\prime|\rightarrow\infty$.
In this case,
$x={|\mathrm{\textbf{r}}-\mathrm{\textbf{r}^\prime}|}{(\xi\hslash)}^{-1}$,
defined by Eq.~(\ref{xxx}), also approaches infinity. Recalling
Eq.~(\ref{greeninh}),
$G(\mathrm{\textbf{r}},t;\mathrm{\textbf{r}}^\prime,t^\prime)$ can
be rewritten in terms of $x$ as
\begin{equation}
\begin{split}
G(\mathrm{\textbf{r}},t;\mathrm{\textbf{r}}^\prime,t^\prime)
=\frac{N\pi^2}{|\mathrm{\textbf{r}}-\mathrm{\textbf{r}^\prime}|}\cdot
\left[H_1\left(x^{\nu}\right)
-iH_2\left(x^{\nu}\right)\right],
\end{split}\label{greeninx}
\end{equation} in which the two H-functions $H_1(x)$ and $H_1(x)$
have been defined by Eqs.~(\ref{hfun1}) and (\ref{hfun2})
respectively.  It is easy to verify that the two H-functions
$H_1(x)$ and $H_1(x)$ satisfy the conditions needed by Property 4 in
the Appendix. Thus, $\mbox{for large }x$, using the asymptotic
formula~(\ref{hproperty4}),  we obtain
\begin{align}
&H_1(x)=\frac{1}{\pi^2}\Gamma(2-\gamma)\sin(\gamma\pi/2)x^{\frac{\gamma-2}{\nu}}+
\frac{M}{\pi i}\cos[\frac{\pi}{4}+(\nu-1)(\frac{x}{\nu^\nu})^{1/(\nu-1)}]x^{\frac{4-2\gamma-\nu}{2\nu(\nu-1)}}+\mbox{o}\left(x^{\frac{4-2\gamma-\nu}{2\nu(\nu-1)}}\right),\label{asmph1}
\\&H_2(x)=\frac{1}{\pi^2}\Gamma(2-\gamma)\sin(\gamma\pi/2)x^{\frac{\gamma-2}{\nu}}-
\frac{M}{\pi i}\sin[\frac{\pi}{4}+(\nu-1)(\frac{x}{\nu^\nu})^{1/(\nu-1)}]x^{\frac{4-2\gamma-\nu}{2\nu(\nu-1)}}+\mbox{o}\left(x^{\frac{4-2\gamma-\nu}{2\nu(\nu-1)}}\right),\label{asmph2}
\end{align}
in which $$M=\frac{\nu^{(2\gamma-3)/[2(\nu-1)]}}{\sqrt{2\pi(\nu-1)}}.$$
Then, from
Eq.~(\ref{greeninh}), we get the asymptotic formula of
$G(\mathrm{\textbf{r}},t;\mathrm{\textbf{r}}^\prime,t^\prime)$ for
large $|\mathrm{\textbf{r}}-\mathrm{\textbf{r}^\prime}|$,
\begin{equation}
\begin{split}
G(\mathrm{\textbf{r}},t;\mathrm{\textbf{r}}^\prime,t^\prime)=\frac{A\mbox{e}^{i\pi/4}|\mathrm{\textbf{r}}-\mathrm{\textbf{r}^\prime}|^{\nu-1}}{2\xi^{2+\nu}\hslash^{3+\nu}i}
\exp\left[i(\alpha-1)\left(\frac{|\mathrm{\textbf{r}}-\mathrm{\textbf{r}^\prime}|}{\alpha\xi\hslash}\right)^{\alpha/(\alpha-1)}\right]
+\mbox{o}\left[\left(\frac{|\mathrm{\textbf{r}}-\mathrm{\textbf{r}^\prime}|}{\xi\hslash}\right)^{\nu}\right].
\end{split}\label{asmpforg}
\end{equation} \\When $\alpha=2$, $\beta=1$, after some calculations, Eq.~(\ref{asmpforg}) reduces to
\begin{equation}
\begin{split}
G(\mathrm{\textbf{r}},t;\mathrm{\textbf{r}}^\prime,t^\prime)=\frac{1}{i\hslash}\left(\frac{m}{2\pi\hslash(t-t^\prime)i}\right)^{3/2}
\exp\left(\frac{im|\mathrm{\textbf{r}}-\mathrm{\textbf{r}^\prime}|^2}{2\hslash(t-t^\prime)}\right)
+\mbox{o}\left(\frac{\sqrt{2m}|\mathrm{\textbf{r}}-\mathrm{\textbf{r}^\prime}|}{\sqrt{(t-t^\prime)\hslash}}\right),
\end{split}\label{asmpforg2}
\end{equation} which accords with the exact result in the standard quantum mechanics \cite{Economou}.
\par
In the scattering problem, we can use merely the first term of the
asymptotic formula (\ref{asmpforg}) for
$G(\mathrm{\textbf{r}},t;\mathrm{\textbf{r}}^\prime,t^\prime)$, then
an approximate wave function for the scattering problem can be
obtained. Let's invoke the Born approximation \cite{grif,born}:
Suppose the incoming plane wave is not substantially altered by the
potential. Then, in Eq.~(\ref{integralform}), it makes sense to use
\begin{equation}
\psi(\mathrm{\textbf{r}}^\prime,t^\prime)\approx\psi_0(\mathrm{\textbf{r}}^\prime,t^\prime)=\mbox{e}^{i(\mathrm{\textbf{k}}\cdot\mathrm{\textbf{r}}^\prime-Et^\prime)/\hslash}.\label{origin2}
\end{equation} Then Eq.~(\ref{integralform}) becomes
\begin{equation}
\begin{split}
\psi(\mathrm{\textbf{r}},t)\approx\psi_0&(\mathrm{\textbf{r}},t)+\frac{A\mbox{e}^{i\pi/4}}{2\hslash^{3+\nu}i}\iint
\frac{|\mathrm{\textbf{r}}-\mathrm{\textbf{r}^\prime}|^{\nu-1}}{\xi^{2+\nu}}
\exp\Bigg[i(\alpha-1)\left(\frac{|\mathrm{\textbf{r}}-\mathrm{\textbf{r}^\prime}|}{\alpha\xi\hslash}\right)^{\alpha/(\alpha-1)}
+\\&i(\mathrm{\textbf{k}}\cdot\mathrm{\textbf{r}}^\prime-Et^\prime)/\hslash\Bigg]
V(\mathrm{\textbf{r}}^\prime,t^\prime)\mathrm{d}^3\mathrm{\textbf{r}}^\prime\mathrm{d}t^\prime.\label{approxwave1}
\end{split}
\end{equation}Furthermore, considering $|\mathrm{\textbf{r}}-\mathrm{\textbf{r}^\prime}|\approx|\mathrm{\textbf{r}}|=r$, we can get
\begin{equation}
\begin{split}
\psi(\mathrm{\textbf{r}},t)\approx\psi_0&(\mathrm{\textbf{r}},t)+\frac{A\mbox{e}^{i\pi/4}r^{\nu-1}}{2\hslash^{3+\nu}i}\iint
\frac{1}{\xi^{2+\nu}}
\exp\Bigg[i(\alpha-1)\left(\frac{r}{\alpha\xi\hslash}\right)^{\alpha/(\alpha-1)}
+i(\mathrm{\textbf{k}}\cdot\mathrm{\textbf{r}}^\prime-Et^\prime)/\hslash\Bigg]
V(\mathrm{\textbf{r}}^\prime,t^\prime)\mathrm{d}^3\mathrm{\textbf{r}}^\prime\mathrm{d}t^\prime.\label{approxwave2}
\end{split}
\end{equation}The second term of the right side of the above formula gives the
approximate scattering wave function.
\par We can also generate a
series of higher-order corrections to the approximate wave function.
From Eq.~(\ref{integralform}), we can build an iteration scheme for
the wave function as
\begin{equation}
\psi^{(n)}(\mathrm{\textbf{r}},t)=\psi_0(\mathrm{\textbf{r}},t)+\iint
G(\mathrm{\textbf{r}},t;\mathrm{\textbf{r}}^\prime,t^\prime)V(\mathrm{\textbf{r}}^\prime,t^\prime)\psi^{(n-1)}(\mathrm{\textbf{r}}^\prime,t^\prime)
\mathrm{d}^3\mathrm{\textbf{r}}^\prime\mathrm{d}t^\prime.\label{iteration}
\end{equation}Note that
$\phi^{(0)}(\mathrm{\textbf{r}},t)=\psi_0(\mathrm{\textbf{r}},t)=\mbox{e}^{i(\mathrm{\textbf{k}}\cdot\mathrm{\textbf{r}}-Et)/\hslash}$,
and $\phi^{(n)}$ is the $n$th-order corrections to the wave
function. For a given potential function, using
Eq.~(\ref{iteration}), the analytical approximate solutions of every
order can be obtained. In a series form, we have
\begin{equation}
\phi=\phi_0+\int GV\phi_0+\iint GVGV\phi_0+\iiint
GVGVGV\phi_0+\cdots.\label{bornseries}
\end{equation}In each integrand only the incident wave function ($\phi_0$)
appears, together with more and more powers of $GV$.

\section{\uppercase{Conclusions}}
In this paper, the three-dimensional space-time-fractional Schr\"odinger equation with time-dependent potential is
studied. We define the Green's function of the STFSE for the
time-dependent scattering problem in the fractional quantum
mechanics, and the STFSE is converted into an integral form. We give
the mathematical expression of the Green's function in terms of
Fox's H-function  and in a computable series form. The asymptotic
property of the Green's function for
$|\mathrm{\textbf{r}}-\mathrm{\textbf{r}}^\prime|\rightarrow\infty$
(or $|\mathrm{\textbf{r}}|>>|\mathrm{\textbf{r}}^\prime|$) was also
given. In this case, the Green's function acts like an exponential
function (see Eq.~(\ref{asmpforg})). Using this result, We obtained
the approximate scattering wave function for the time-dependent
fractional quantum scattering problem (see Eq.~(\ref{approxwave2})).
A series of higher-order corrections to the approximate wave
function were also given in Eq.~(\ref{bornseries}). These results are
useful for the time-dependent scattering problem in the fractional
quantum mechanics. All of these results contain those in the
standard quantum mechanics as special cases.

\section*{{Acknowledgements}}
This work was supported by the National Natural Science Foundation of China (Grant
No. 11147109), the Specialized Research Fund for the Doctoral Program of Higher Education
of China (Grant No. 20113218120030), and the Fundamental Research Funds for the Central
Universities (Grant No. NS2012119).

\appendix
\renewcommand{\theequation}{A\arabic{equation}}
\setcounter{equation}{0}
\renewcommand{\thesection}{APPENDIX:}
\section{\uppercase{Fox's $H$-function and Some Properties}}
The Fox's $H$-function \cite{mathai0,mathai} is defined by an
integral of Mellin-Barnes type \cite{Paris} as
\begin{equation}H^{m,n}_{p,q}(z)=H^{m,n}_{p,q}\left[z\text{ }\Bigg|
\begin{aligned}&{\text{
}(a_1,A_1),(a_2,A_2),\ldots,(a_p,A_p)}
\\&{\text{
}(b_1,B_1),(b_2,B_2),\ldots,(b_p,B_p)}
\end{aligned}\right]=\cfrac1{2\pi
i}\int _{L}\chi(s)z^{-s}\mathrm{d}s,\label{a1}\end{equation}where
\begin{equation}
\chi(s)=\cfrac{\prod^m_{j=1}\Gamma(b_j+B_js)\prod^n_{i=1}\Gamma(1-a_i-A_is)}{\prod^p_{i=n+1}\Gamma(a_i+A_is)\prod^q_{j=m+1}\Gamma(1-b_j-B_js)}.\label{a1add}
\end{equation} The contour $L$ runs from $c-i\infty$ to $c+i\infty$ separating the
poles of $\Gamma(1-a_i-A_is)$, ($i=1,\cdots,n$) from those of
$\Gamma(b_j+B_js)$, ($j=1,\cdots,m$).  Here we present some
properties of the $H$-function used in our paper. In order to give
the results, the following definitions will be used,
\begin{equation}
\begin{aligned}
&\Delta=\sum^q_{j=1}B_j-\sum^p_{i=1}A_j;\quad\Delta^*=\sum^n_{i=1}A_i-\sum^p_{i=n+1}A_i+\sum^m_{j=1}B_j-\sum^q_{j=m+1}B_j;
\\&\delta=\prod^p_{j=1}(A_j)^{-A_j}\prod^q_{j=1}(B_j)^{B_j};
\quad\mu=\sum^q_{j=1}b_j-\sum^p_{i=1}a_i+\frac{p-q}{2}.
\end{aligned}\label{a2}
\end{equation}The following properties of the  $H$-function can be
found in Refs.~\cite{mathai,mathai0,kilbas0}.
\\\textbf{Property 1:}\\
\begin{align}
\frac{1}{k} H^{m,n}_{p,q}\left[z\text{ }\Bigg|
\begin{aligned}&{\text{
}(a_p,A_p)}
\\&{\text{
}(b_p,B_p))}
\end{aligned}\right]=H^{m,n}_{p,q}\left[z^k\text{ }\Bigg|
\begin{aligned}&{\text{
}(a_p,kA_p)}
\\&{\text{
}(b_p,kB_p)}
\end{aligned}\right].\quad\mbox{for } k>0\label{hproperty1}
\end{align}
\textbf{Property 2:}\\
\begin{align}
z^\sigma H^{m,n}_{p,q}\left[z\text{ }\Bigg|
\begin{aligned}&{\text{
}(a_p,A_p)}
\\&{\text{
}(b_p,B_p))}
\end{aligned}\right]=H^{m,n}_{p,q}\left[z\text{ }\Bigg|
\begin{aligned}&{\text{
}(a_p+\sigma A_p,A_p)}
\\&{\text{
}(b_p+\sigma B_p,B_p)}
\end{aligned}\right].\quad\mbox{for } \sigma\in\mathbb{C}\label{hproperty2}
\end{align}
\textbf{Property 3: Explicit Power Series Expansion }\par For
$\Delta>0$ , $z\neq0$ or $\Delta=0$, $|z|>\delta$, there holds the
following expansion for the $H$-function \cite{mathai},
\begin{align}
H^{m,n}_{p,q}(z)=\sum^{m}_{h=1}\sum^{\infty}_{k=0}\cfrac{\prod^m_{j=1,j\neq
h}\Gamma(b_j-B_js_{hk})\prod^n_{i=1}\Gamma(1-a_i+A_is_{hk})}
{\prod^p_{i=n+1}\Gamma(a_i-A_is_{hk})\prod^q_{j=m+1}\Gamma(1-b_j+B_js_{hk})}\frac{(-1)^k}{k!}\frac{z^{s_{hk}}}{B_h},\label{hproperty3}
\end{align} where $s_{hk}=(b_h+k)/B_h$,
if the following conditions are satisfied:
\begin{enumerate}
\item The poles of the gamma functions $\Gamma(1-a_i-A_is)$, ($i=1,\cdots,n$) and those
of $\Gamma(b_j+B_js)$, ($j=1,\cdots,m$) do not coincide:
\begin{equation}
A_i(b_j+l)\neq B_j(a_i-k-1), (i=1,\cdots,n;
j=1,\cdots,m;k,l=0,1,2,\cdots).\label{condition1}
\end{equation}
\item The poles of the gamma functions $\Gamma(b_j+B_js)$,
($j=1,\cdots,m$) are simple:
\begin{equation}
B_i(b_j+l)\neq B_j(b_i+k), (i\neq
j;i,j=1,\cdots,m;k,l=0,1,2,\cdots).\label{condition2}
\end{equation}
\end{enumerate}
\ \textbf{Property 4: Asymptotic Expansions at Infinity in the Case
$\Delta>0$, $\Delta^*=0$ }\par When the poles of the gamma functions
$\Gamma(1-a_i-A_is)$, ($i=1,\cdots,n$) are simple:
\begin{equation} A_i(1-A_j+l)\neq
A_j(1-a_i+k), (i\neq
j;i,j=1,\cdots,n;k,l=0,1,2,\cdots),\label{condition3}
\end{equation} and the condition (\ref{condition1})
is satisfied, the $H$-function has the following asymptotic
expansion \cite{kilbas,kilbas0},
\begin{equation}
\begin{split}
H^{m,n}_{p,q}(z)=&\sum^{n}_{i=1}\left[h_iz^{(a_i-1)/A_i}+\mbox{o}\left(z^{(a_i-1)/A_i}\right)\right]+Az^{(\mu+1/2)/\Delta}\Big(c_0\exp\left[(B+Cz^{1/\Delta})i\right]-
\\&d_0\exp\left[-(B+Cz^{1/\Delta})i\right]\Big)+\mbox{o}\left(z^{(\mu+1/2)/\Delta}\right),\label{hproperty4}
\end{split}
\end{equation}
where \begin{align}
&h_i=\frac{1}{A_i}\cfrac{\prod^m_{j=1}\Gamma(b_j-B_j(a_i-1)/A_i)\prod^n_{j=1,j\neq
i}\Gamma(1-a_j+A_j(a_i-1)/A_i)}
{\prod^p_{j=n+1}\Gamma(a_j-A_j(a_i-1)/A_i)\prod^q_{j=m+1}\Gamma(1-b_j+B_j(a_i-1)/A_i)},\label{hp4hi}
\\&A=\frac{A_0}{2\pi
i\Delta}\left(\frac{\Delta^\Delta}{\delta}\right)^{(\mu+1/2)/\Delta},\quad
B=\frac{(2\mu+1)\pi}{4}, \quad
C=\left(\frac{\Delta^\Delta}{\delta}\right)^{1/\Delta},\label{hp4abc}
\\&A_0=(2\pi)^{(p-q+1)/2}\Delta^{-\mu}\prod^p_{i=1}A_i^{-a_i+1/2}\prod^q_{j=1}B_j^{b_j-1/2},\label{hp4a0}
\\&c_0=(2\pi
i)^{m+n-p}\exp\left[\left(\sum^p_{i=n+1}a_i-\sum^m_{j=1}b_j\right)\pi
i\right],\label{hp4c0}
\\\mbox{and } \quad &d_0=(-2\pi
i)^{m+n-p}\exp\left[-\left(\sum^p_{i=n+1}a_i-\sum^m_{j=1}b_j\right)\pi
i\right].\label{hp4d0} \end{align}

\end{document}